\documentclass[aps,prd, 10pt, twocolumn,nofootinbib,showpacs]{revtex4-1}
\usepackage{subeqn}
\usepackage[usenames,dvipsnames]{xcolor}

\begin{document}

\title{Is there a mode stability paradox for neutrino perturbations of Kerr black holes?}
\author{Koray D\"{u}zta\c{s}}
\email{duztasko@hotmail.com}
\affiliation{Bo\u{g}azi\c{c}i University, Department of Physics \\ Bebek 34342, \.Istanbul, Turkey}

\begin{abstract}
Adopting the notation of Teukolsky and Press, we derive the connection relation for asymptotic solutions of massless Dirac equation on Kerr background. We show that, unlike bosonic fields, the connection relation for massless Dirac fields (neutrino) provides a rigorous proof of mode stability. The same relation also implies that that every incoming mode can be absorbed by the black hole or there is no superradiance. Recent works on overspinning black holes have shown that this can lead to formation of naked singularities. We argue that, the fact that both the mode stability of the black hole under neutrino perturbations, and the instability of the event horizon (therefore the instability of the black hole) can be derived from the same connection relation, leads to a paradox. In other words mode stability implies event horizon instability as far as neutrino perturbations are concerned. 
\end{abstract}
\maketitle
\section{Introduction}
Black hole solutions should be accompanied with a proof stability so that they can describe actual astrophysical objects in the universe. If a known solution turns out to be unstable it may correspond to an intermediate stage of gravitational collapse. In that case the final configuration remains unknown. For example, an unstable Kerr black hole might radiate away mass and angular momentum to settle down to a stable Kerr solution, or the final state could be a non-Kerr, stationary, axi-symmetric configuration with an event horizon, provided that such a solution exists. In the worst case scenario, the hypothesized unstable black hole evolves to a naked singularity, which is visible from asymptotically flat infinity.

In the classical mode analysis for stability, one studies the individual modes of gravitational perturbations of a black hole space-time, to check if there exists any modes that grow without bound in time. The same analysis can also be applied to other massless fields of interest. The existence of exponentially growing modes proves that the black hole is unstable, since it will not be able to return to its original state after perturbation. 

In this context, the stability of Schwarzschild solution  was proved much earlier than Kerr solution, which is more convenient to describe astrophysical objects with its additional angular momentum parameter~\cite{sch1,sch2,sch3,sch4}. The fact that the energy density is not positive in  the ergosphere, and the effective potential associated with the radial part of the Teukolsky equation~\cite{teuk1} --which describes all massless fields on Kerr background-- makes it hard to give a rigorous proof of stability analogous to Schwarzschild case. Teukolsky and Press found a conserved energy for the radial equation~\cite{teuk2,teuk3}, and searched for instabilities by numerical integration of perturbation equations.  Finally Whiting applied differential and integral transformations of Kerr radial and angular functions to construct a conserved quantity with a positive definite integrand; hence gave a rigorous proof that no exponentially growing modes of Kerr solution exists~\cite{whiting}. Whiting's result can be considered as the culmination of the classical mode analysis of the stability problem. (Also see \cite{teukrev})

Mode stability does not necessarily imply that a general linear perturbation with a finite initial energy cannot grow  in time without bound. To reach that conclusion one should give a proof of linear stability beyond mode analysis. This was accomplished by Kay and Wald for Schwarzschild black holes~\cite{kaywald}, but a rigorous proof has not been carried out for the Kerr case.  (See \cite{daf1,daf2,daf3}). Stability problem was evaluated for several black hole solutions in four and higher dimensions~\cite{ref1,ref2,ref3,ref4,ref5,ref6,ref7,ref8,ref9,ref10,ref11,ref12,ref13,ref14,ref15}

In this work we re-visit the stability problem of Kerr black holes under massless Dirac perturbations. First, we follow the recipe developed by Teukolsky and Press to derive the   connection relation for the asymptotic coefficients of massless Dirac equation on Kerr background. Using this connection relation, we give a rigorous proof of mode stability for neutrino perturbations, without referring to Whiting's result. Based on recent works on overspinning black holes, we also show that one can infer the instability of the event horizon from the same connection relation. We argue that this leads to a paradox.
\subsection{Dirac equation in Kerr space-time}
\label{sec:1}
The two-spinor formalism developed by Newman and Penrose (NP)~\cite{newpen} has been very useful to evaluate field equations in curved space-times. Dirac equation couples two valence 1 spinors corresponding to spin 1/2 fields

\begin{eqnarray}
& &\nabla_{A \dot{A}}P^{A} + i \mu_f \bar{Q}_{\dot{A}}=0 \nonumber \\
& & \nabla_{A \dot{A}}Q^{A} + i \mu_f \bar{P}_{\dot{A}}=0   \label{diraceqn} 
\end{eqnarray}
where $\nabla_{A \dot{A}}$ is the spinor covariant derivative (see \cite{penrosebook,agr}) and $\sqrt{2}\mu_f$ is the mass of the fermion field. A tetrad of null vectors is induced by the spin basis.
\begin{equation}
l^a=o^A \bar{o}^{A'} \quad n^a=\iota^A \bar{\iota}^{A'} \quad m^a=o^A \bar{\iota}^{A'} \quad \bar{m}^a=\iota^A \bar{o}^{A'} \label{nptetrad}
\end{equation}
$l$ and $n$ are real while $m$ and $\bar{m}$ are complex conjugates of each other. Null vectors satisfy orthogonality relations
\begin{eqnarray}
& &l_an^a=n_al^a=-m_a \bar{m}^a=-\bar{m}_a m^a=1 \nonumber \\
& &l_am^a=l_a \bar{m}^a=n_am^a=n_a\bar{m}^a=0
\end{eqnarray}
The directional derivatives along the null directions are denoted by conventional symbols
\begin{equation}
D=l^a\nabla_a,\quad \Delta=n^a\nabla_a,\quad \delta=m^a\nabla_a,\quad \bar{\delta}=\bar{m}^a\nabla_a
\end{equation}
One can express $\nabla_a$ in terms of NP derivative operators:
\begin{eqnarray}
\nabla_a &=&g_a^{\;\;b}\nabla_b \nonumber \\
&=& (n_al^b+l_an^b-\bar{m}_am^b-m_a\bar{m}^b)\nabla_b  \nonumber \\
&=& n_aD + l_a \Delta - \bar{m}_a \delta -m_a \bar{\delta} \label{nablanp}
\end{eqnarray}
The explicit form of  Dirac's equations (\ref{diraceqn}) in NP formalism is~\cite{chandrabook}
\begin{eqnarray}
& &(D+\epsilon-\rho)P^0+ (\bar{\delta}+\pi -\alpha)P^1=i\mu_f \bar{Q}^{\dot{1}} \label{dirac1} \\
& &(\Delta+\mu -\gamma )P^1 + (\delta -\tau +\beta )P^0=-i\mu_f \bar{Q}^{\dot{0}} \label{dirac2} \\
& & (D+ \bar{\epsilon} -\bar{\rho} )\bar{Q}^{\dot{0}}+ (\delta +\bar{\pi} -\bar{\alpha})\bar{Q}^{\dot{1}}=-i\mu_f P^1 \label{dirac3} \\
& & (\Delta +\bar{\mu} - \bar{\gamma})\bar{Q}^{\dot{1}} +(\bar{\delta} + \bar{\beta} - \bar{\tau})\bar{Q}^{\dot{0}}=i \mu_f P^0 \label{dirac4}
\end{eqnarray}
where $P^0,Q^O$ and $P^1,Q^1$ are components of $P^A,Q^A$ along the spinor dyad basis $o^A$ and $\iota^A$ respectively. Chandrasekhar \cite{chandradirac} showed that Dirac equation is separable on Kerr background.
\begin{eqnarray}
P^0 &=& (r-ia\cos \theta)^{-1}[{}_{-1/2}R(r)][{}_{-1/2}S(\theta)] e^{-i\omega t }e^{i m\phi} \nonumber \\
P^1 &=& [{}_{1/2}R(r)][{}_{1/2}S(\theta)] e^{-i\omega t }e^{i m\phi} \nonumber \\
Q^{\dot{0}} &=&  -(r+i a\cos \theta)^{-1}[{}_{-1/2}R(r)][{}_{1/2}S(\theta)] e^{-i\omega t }e^{im\phi} \nonumber \\
Q^{\dot{1}} &=&  [{}_{1/2}R(r)][{}_{-1/2}S(\theta)] e^{-i\omega t }e^{i m\phi}
\label{diracsep}
\end{eqnarray}
Letting $\mu_f =0$ for the massless case, Chandrasekhar's separation leads to a pair of equations for both $ [{}_{\pm1/2}S(\theta)]$ and $[{}_{\pm1/2}R(r)]$. The angular equation takes the form
\begin{eqnarray}
&&\frac{1}{\sin \theta}\frac{\partial}{\partial \theta} \left( \sin \theta \frac{\partial S}{\partial \theta} \right) 
 +\left( a^2\omega^2 \cos^2 \theta -\frac{m^2}{\sin^2 \theta}-2a\omega s \cos \theta \right. \nonumber \\
&&\left. - \frac{2ms\cos \theta}{\sin^2 \theta}-s^2 \cot^2 \theta +E-s^2 \right)S=0 
\label{teukangular}
\end{eqnarray}
where $E$ is the eigenvalue. The radial functions satisfy
\begin{eqnarray}
&&\Delta^{-s} \frac{\partial}{\partial r} \left(\Delta^{s+1} \frac{\partial ({}_sR)}{\partial r} \right) \nonumber \\
&&+ \left( \frac{K^2 -2i s(r-M)K}{\Delta} +4i s \omega r -\lambda \right)({}_sR)=0
\label{teukradial}
\end{eqnarray}
where $K \equiv (r^2+a^2)\omega - am$ and $\lambda \equiv E + a^2 \omega^2 -2am\omega - s(s+1)$. 

As we let $\mu=0$ in Chandrasekhar's separation we have derived  Teukolsky's equations for massless fields \cite{teuk1}. The asymptotic solutions at infinity for the radial functions are 
\begin{eqnarray}
{}_{-1/2}R&=& Z_{\rm{in}}\frac{e^{-i\omega r*}}{r}+ Z_{\rm{out}}e^{i \omega r*} \nonumber \\
{}_{1/2}R&=& Y_{\rm{in}}\frac{e^{-i\omega r*}}{r}+ Y_{\rm{out}}\frac{e^{i \omega r*}}{r^2} 
\label{radialsolnsinf}
\end{eqnarray}
We have adopted the notation of Teukolsky and Press \cite{teuk3}. $Y_{\rm{in}}$, $Y_{\rm{out}}$, $Z_{\rm{in}}$, $Z_{\rm{out}}$ are the amplitudes of the ingoing and outgoing waves at infinity for the cases $s=1/2$ and $s=-1/2$ respectively. We are going to derive a connection relation between these coefficients, which represents a conservation law relating the net absorption of the wave at infinity to the net increase in the mass parameter of Kerr space-time. $r^*$ is the tortoise coordinate defined by $dr^*/dr=(r^2+a^2)/\Delta$. The asymptotic solutions of the Dirac equation are evaluated at $r^* \to -\infty$ $(r\to \infty)$, and $r^* \to -\infty$  $(r \to r_+)$, where $r_+$ denotes the spatial coordinate of the event horizon. 

Near the horizon, only the ingoing solutions of the radial equation (\ref{teukradial})  are physical
\begin{eqnarray}
 && \left. Y_{\rm{hole}}(\Delta^{-1/2}) e^{-i kr*} \right\} s=1/2 \nonumber \\
 & &\left.  Z_{\rm{hole}}(\Delta^{1/2})  e^{-i kr*}  \right\} s=-1/2
\label{teuksolnshor}
\end{eqnarray}
where $\Omega=a/(2Mr_+)$ is the rotational frequency of the black hole, and $k=\omega-m\Omega$.

\section{Connection relations}
We follow the methods of Teukolsly and Press \cite{teuk3} to derive the connection relations between the asymptotic solutions at infinity and at the horizon for neutrino fields. For the angular part, the equation (\ref{dirac1}) is separated to give
\begin{equation}
-\mathcal{L}[{}_{1/2}S]=B[{}_{-1/2}S] \label{angdirac1}
\end{equation}
where 
\begin{equation}
\mathcal{L}=\partial_{\theta}+\frac{m}{\sin \theta}-a\omega \sin \theta +  \frac{\cot \theta}{2}
\end{equation}
$\mathcal{L}$ agrees with the angular operator defined in \cite{teuk3} for $n=1/2$. Note that $B$ is real since the differential operator  $\mathcal{L}$ and the angular functions $S$ are real. The fact that $B$ is real will be crucial for the proof of mode stability in the next section. For the radial part (\ref{dirac1})  gives
\begin{equation}
\left(\partial_r -\frac{iK}{\Delta}\right)[{}_{-1/2}R]=\frac{B}{\sqrt{2}}[{}_{1/2}R]
\label{radialdirac1}
\end{equation}
The asymptotic solutions at infinity for radial functions were given in  (\ref{radialsolnsinf}). Now we substitute these solutions in (\ref{radialdirac1}) and evaluate at $r\to \infty$. This leads to the connection between $ Y_{\rm{in}}$ and  $ Z_{\rm{in}}$.
\begin{equation}
B Y_{\rm{in}}=-2\sqrt{2}i\omega Z_{\rm{in}}
\label{yinzin}
\end{equation}
Similarly (\ref{dirac2}) gives
\begin{equation}
\mathcal{L}^{\dagger}[{}_{-1/2}S]=B[{}_{1/2}S] \label{angdirac2}
\end{equation}
where $\mathcal{L}^{\dagger}=\mathcal{L}(-\omega ,-m)$. (\ref{angdirac1}) and (\ref{angdirac2}) imply
\begin{eqnarray}
&&(-\mathcal{L})(\mathcal{L}^{\dagger})[{}_{-1/2}S]=B^2[{}_{-1/2}S] \label{b1} \\
&&(\mathcal{L}^{\dagger})(-\mathcal{L})[{}_{1/2}S]=B^2[{}_{1/2}S] \label{b2}
\end{eqnarray}
The angular functions $[{}_{\pm1/2}S]$ satisfy Teukolsky's angular equation (\ref{teukangular}) for $s=\pm 1/2$. Comparing (\ref{b1}) and (\ref{b2}) with (\ref{teukangular}) one may also express $B$ in terms of $(a,m,\omega)$.
\begin{equation}
B^2=E+a^2\omega^2 - 2a\omega m + 1/4
\end{equation}
We proceed to evaluate (\ref{dirac2}) for the radial part.
\begin{equation}
\Delta\left( \partial_r +\frac{i K}{\Delta}+\frac{r-M}{\Delta}\right)[{}_{1/2}R]=B\sqrt{2}[{}_{-1/2}R]
\label{radialdirac2}
\end{equation}
Substituting the asymptotic solutions at infinity for the radial functions in (\ref{radialdirac2}), and evaluating at $r\to \infty$, one derives  the connection between $Y_{\rm{out}}$ and $Z_{\rm{out}}$.
\begin{equation}
\sqrt{2}i\omega Y_{\rm{out}}=BZ_{\rm{out}}
\label{youtzout}
\end{equation} 
To find the connection relation between $Y_{\rm{hole}}$ and $Z_{\rm{hole}}$, we  evaluate (\ref{radialdirac1}) at $r\to r_+$, using the solutions of the radial equation near the horizon.
\begin{equation}
B Y_{\rm{hole}}=\sqrt{2}\{ (r_+ -M)-4i kMr_+\}Z_{\rm{hole}} \label{yholezhole}
\end{equation}
The form of the radial differential equation (\ref{teukradial}) allows us to find a connection relation between the asymptotic solutions at infinity and at the horizon. For that purpose we follow Teukolsky and Press \cite{teuk3} to write the radial Teukolsky equation in the form:
\begin{equation}
X_{,r^*r^*}+VX=0  \label{Yeqn}
\end{equation}
where
\begin{equation}
X(s)=\Delta^{s/2}(r^2+a^2)^{1/2}({}_sR) 
\end{equation} 
The potential $V$ is given by
\begin{eqnarray}
&&V= [K^2-2isK(r-M) \nonumber \\
&&+ \Delta(4i rws-Q)-s^2(M^2-a^2)]/(r^2+a^2)^2 \nonumber \\
& &-\Delta[2Mr^3 +a^2r^2 -4Mra^2+a^4]/(r^2+a^2)^4 \label{V1}
\end{eqnarray}
where $K=(r^2 + a^2)w - am$ and $Q=E + a^2w^2 -2awm$. The Wronskian of any two solutions of (\ref{Yeqn}) is conserved. Since the potential satisfies $V(r,w,m,l,s,a)=V^*(r,w,m,l,-s,a)$, two linearly independent solutions of (\ref{Yeqn}) are $X(s)$ and $X^*(-s)$. Thus
\begin{equation}
[X_{+,r^*}X^{*}_--X_+X^{*}_{-,r^*}]_{r_+}=[X_{+,r^*}X^{*}_--X_+X^{*}_{-,r^*}]_{\infty} \label{wron1}
\end{equation}
$X_+$ and $X_-$ in (\ref{wron1}) denote $X(s)$ and $X(-s)$, respectively. Near the horizon the form of $Y(\pm1/2)$ is
\begin{eqnarray}
&&X(1/2)_{\rm{H}}= Y_{\rm{hole}}\Delta^{-1/4}(r^2+a^2)^{1/2}e^{-i kr*} \nonumber \\
&&X(-1/2)_{\rm{H}}= Z_{\rm{hole}}\Delta^{1/4}(r^2+a^2)^{1/2}e^{-i kr*}
\end{eqnarray}
The subscript ``H'' refers to the horizon. Let us evaluate the left hand side of (\ref{wron1}).
\begin{eqnarray}
\rm{LHS}&=&Y_{\rm{hole}}Z^*_{\rm{hole}}[-(r_+-M)-2i k(r^2+a^2)] \nonumber \\
&=& -\frac{B\vert Y_{\rm{hole}}\vert^2}{\sqrt{2}} \label{lhs}
\end{eqnarray}
where we used (\ref{yholezhole}). The form of $X(\pm1/2)$ at infinity is
\begin{eqnarray}
&&X_+= \Delta^{1/4}(r^2+a^2)^{1/2}\left(Y_{\rm{in}}\frac{e^{-i\omega r*}}{r}+Y_{\rm{out}}\frac{e^{i\omega r*}}{r^2} \right)\nonumber \\
&&X_-= \Delta^{-1/4}(r^2+a^2)^{1/2}\left(Z_{\rm{in}}\frac{e^{-i\omega r*}}{r}+Z_{\rm{out}}e^{i\omega r*} \right) 
\end{eqnarray}
Then the right hand side of (\ref{wron1}) becomes
\begin{eqnarray}
\rm{RHS}&=&2i\omega (-Y_{\rm{in}}Z^*_{\rm{in}}+Y_{\rm{out}}Z^*_{\rm{out}}) \nonumber \\
&=&-\frac{B\vert Y_{\rm{in}}\vert^2}{\sqrt{2}} +B\sqrt{2}\vert Z_{\rm{out}} \vert^2
\label{rhs}
\end{eqnarray}
where we used (\ref{yinzin}) and (\ref{youtzout}).  Equating the two sides of (\ref{wron1}) we derive the connection relation between the asymptotic solutions for massless Dirac equation in Kerr background.
\begin{equation}
\frac{\vert Y_{\rm{in}}\vert^2}{2}-\vert Z_{\rm{out}} \vert^2=\frac{\vert Y_{\rm{hole}}\vert^2}{2} \label{conneut}
\end{equation}
The corresponding connection relations for scalar $(s=0)$, electromagnetic $(s=1)$ and gravitational $(s=2)$ perturbations were derived in \cite{teuk3}
\begin{equation}
\left. \vert Z_{\rm{in}} \vert^2-\vert Z_{\rm{out}} \vert^2=2Mr_+(\frac{k}{\omega})\vert Z_{\rm{hole}} \vert^2 \right \} s=0 \label{conscalar}
\end{equation}
\begin{equation}
 \left. \frac{\vert Y_{\rm{in}}\vert^2}{4}-\vert Z_{\rm{out}} \vert^2=\frac{\omega}{4k(2Mr_+)}\vert Y_{\rm{hole}}\vert^2 \right \} s=1 \label{conem}
\end{equation}
\begin{equation}
 \left. \frac{\vert Y_{\rm{in}}\vert^2}{16}-\vert Z_{\rm{out}} \vert^2=\frac{\omega^3}{16k(2Mr_+)^3k^{\prime^2}}\vert Y_{\rm{hole}}\vert^2 \right \} s=2 \label{congrav}
\end{equation}
where $k^{\prime^2}=(k^2+4\epsilon^2)$ and $\epsilon=(M^2-a^2)^{1/2}/(4Mr_+)$. Since $(+s)$ solutions are dominant for ingoing waves at infinity (and at the horizon), and $(-s)$ solutions are dominant for outgoing waves at infinity, we expressed the connection relations in terms of $Y_{\rm{in}}$,  $Y_{\rm{hole}}$, and  $Z_{\rm{out}}$. (Teukolsky and Press express these relations in terms of $Y_{\rm{in}}$,  $Y_{\rm{hole}}$, and  $Y_{\rm{out}}$.) These connection relations are also explicit expressions of energy conservation, relating the net absorption of the wave incident from infinity to the net increase in black hole's energy (see \cite{teuk3}).
\begin{equation}
\frac{dE_{\rm{in}}}{dt}-\frac{dE_{\rm{out}}}{dt}=\frac{dE_{\rm{hole}}}{dt}
\end{equation}

\section{Mode stability}
The equations (\ref{teukradial}) and (\ref{teukangular}) which describe perturbations of Kerr black holes, determine a non-linear eigenvalue problem for the frequency $\omega$. The non-linearity arises due to the fact that the eigenvalues $E,\lambda$ explicitly depend on $\omega$. In \cite{teuk2} Press and Teukolsky argued that the problem of mode stability  of these perturbations can be reduced to a rather straightforward evaluation. If there are no solutions of the radial equation with $\omega$ in the upper half complex plane for all angular modes $(l,m)$, then the corresponding perturbations cannot grow without bound and stability is guaranteed. If there are eigen-frequencies in the upper half plane, these correspond to unstable modes which grow exponentially in time. Press and Teukolsky also showed that an instability corresponds to the case that the incoming wave at infinity has zero amplitude for $\omega$ in the upper half plane. To examine the upper half complex $\omega$ plane, one writes the radial solution in the form
\begin{equation}
R=Y_{\rm{in}} e^{-i \omega r*}/r +  Y_{\rm{out}}  e^{i \omega r*}/r^{(2s +1)}
\label{teuk2} 
\end{equation}
Then one seeks for zeros of $Y_{\rm{in}}$ or poles of $Y_{\rm{out}} /Y_{\rm{in}}$, viewed as a function of $\omega$ in the complex plane. The fact that the Schwarzschild solution is  known to be stable simplifies the problem.  For $a=0$ all the zeros of $Y_{\rm{in}}$ must lie in the lower half plane. An instability can only occur if a pole migrates smoothly from the lower half plane and crosses the real axis, as we spin up the black hole from the stable case $a=0$~\cite{hartle}. Therefore the search for instabilities can be restricted to the real axis without loss of generality.

In that respect, the connection relations (or the conserved energy) that were derived from the two linearly independent solutions of the radial equation, can be used  to search for unstable modes. Let us first consider the case of bosonic test fields. When we let $Y_{\rm{in}}=0$ in equations (\ref{conscalar}), (\ref{conem}), and (\ref{congrav}), we see that only in the range $\omega \in (0, m\Omega)$, the two sides of the connection relations for bosonic fields can have the same sign. In other words unstable modes can only exist in this range. This range was examined numerically and no instabilities were found~\cite{teuk2,teuk3}. The connection relations for bosonic fields provided an unsuccessful search for instabilities, rather than a general proof of stability. The general analytic proof was given by Whiting~\cite{whiting}.

Let us express the connection relation for neutrino fields (\ref{conneut}) in terms of $Y_{\rm{out}}$ to search for poles. 
\begin{equation}
\frac{\vert Y_{\rm{in}} \vert^2}{\sqrt{2}} - \frac{2\sqrt{2}\omega^2\vert Y_{\rm{out}} \vert^2}{B^2}=\frac{\vert Y_{\rm{hole}} \vert^2}{\sqrt{2}}
\label{conneut2}
\end{equation}
When $Y_{\rm{in}}=0$, the two sides of the connection relation (\ref{conneut2}) cannot have the same sign for any value of $\omega$, since $B$ is real and $\omega$ can be restricted to the real axis without loss of generality~\cite{hartle}. Thus, no unstable modes exist. This way, the connection relation for neutrino fields provides a rigorous proof of mode stability without referring to Whiting's analysis. 

\section{Overspinning black holes with test bodies, bosonic and fermionic fields}
The stability of the event horizons of black holes is crucial to preserve causality and deterministic nature of general relativity. For that reason the singularities which inevitably ensue as a result of gravitational collapse,  were conjectured to be hidden behind event horizons  by Penrose~\cite{ccc}. It has not been possible to establish a concrete proof of this conjecture, known as cosmic censorship. The studies in this area often remained restricted to case by case analysis of the interactions of black holes with test particles or fields, to test the stability of the event horizon. 

Recently it was shown that a Kerr black hole can be overspun into a naked singularity as a result of the interaction with a test body~\cite{jacobsot} or a bosonic test field~\cite{overspin}. In both cases overspinning occurs  in a narrow frequency range. Later dissipative (radiative) and
conservative (gravitational) self-force effects were considered for the case of test bodies
and it was shown that conservative self-force is comparable to the terms giving rise to naked
singularities~\cite{antij}. In principle, gravitational and radiative self-force effects can also be calculated
for the case of bosonic test  fields to compensate for the overspinning effect, because these two thought experiments are analogous to each other. For a test body that crosses the horizon the flux of energy and angular momentum absorbed by the black hole are related via
\begin{equation}
\delta E - \Omega \delta J= \int T_{ab}\chi^a d\Sigma^b
\end{equation}
where $T_{ab}$ is the energy-momentum tensor, $\chi^a$ is the horizon generating  Killing vector and $\Sigma^b$ is the horizon surface element. The null energy condition implies
\begin{equation}
\delta E > \Omega \delta J
\label{condibody}
\end{equation}
This sets an upper bound on the amount of angular momentum that can be absorbed by the black hole $\delta J < \delta J_{\rm{max}}= (\delta E)/\Omega$. The lower bound for $\delta J$ is found by requiring that the final configuration satisfies $(M+ \delta M)^2- (J + \delta J)<0$ so that cosmic censorship is violated.  Bosonic test fields also satisfy the null energy condition. Each field in the mode $(\omega,l,m)$ represents a large number of particles with energy $\hbar \omega$, and angular momentum $\hbar m$ far away from the black hole; hence $(\delta E)/ (\delta J)=\omega/m$. The condition (\ref{condibody}) can be expressed in the form
\begin{equation}
\omega > \Omega m = \omega_{\rm{sl}}
\end{equation}
For bosonic fields the upper limit for the amount of angular momentum that can be absorbed by the black hole is also the lower limit for the frequency, which corresponds to the superradiance limit $\omega_{\rm{sl}}$.
The absorption of  modes with lower energy and higher angular momentum is not allowed, hence overspinning can be achieved in a limited extent and a narrow range of frequencies.  To be precise, the frequency has to be in the range $\omega_{\rm sl} < \omega < \omega_{1}$, where $\omega_1 \equiv \omega_0 /(1+\sqrt{2}\epsilon )$, $\omega_{0} \equiv m/2M$,  and $\epsilon \ll 1$ is used to parametrise the closeness of the initial black hole to extremality such that $J/M^2=1-2\epsilon^2$. (see \cite{overspin}) Both for test bodies and bosonic test fields the allowed ranges for $\delta J$ and $\delta E$ to violate cosmic censorship is of order $\epsilon^2$ and the final configuration of parameters satisfy $(M+ \delta M)^2- (J + \delta J)\sim - \epsilon^2$.

However it is known that  fermionic fields do not satisfy the null energy condition and therefore superradiance does not occur~\cite{chandrabook}. In this case the waves with frequencies below the superradiance limit can also be absorbed. The range of frequencies that allow overspinning is not bounded below by superradiance; it is extended to $0<\omega<w_1$. (see \cite{superrad}) As $\omega$ is lowered below the superradiant limit, $(\delta J)/(\delta E)$ increases without bound while the final value of $(M+ \delta M)^2- (J + \delta J)$ decreases without bound. (or the absolute value of $(M+ \delta M)^2- (J + \delta J)$  increases without bound.) This is analogous to the case of an imaginary test particle which can violate the condition $\delta E > \Omega \delta J$ to be absorbed by the black hole. For such a test particle the self-force effects would become negligible as $\delta J$ is allowed to grow without bound with respect to $\delta E$, and the formation of a naked singularity could not be avoided.  From that point of view the violation of cosmic censorship by neutrino fields is generic as opposed to similar attempts involving test bodies and bosonic fields. (see~\cite{duztas,superrad,toth,natario})

One should also prove that the  challenging waves are absorbed by the black hole so that the black hole is overspun. Explicit forms of absorption probabilities for massless fields were derived by Page~\cite{page}. For bosonic fields, a factor of $(\omega - m\Omega)$ occurs, which guarantees that the absorption probability goes to zero as the superradiant limit is approached, and it becomes negative for $(\omega < m\Omega)$, i.e. the waves are amplified rather than absorbed. On the other hand the dominant contribution to the absorption probability of a neutrino field is given by
\begin{equation}
\Gamma_{(1/2)\omega(1/2)(1/2)}=M^2\omega^2
\end{equation}
The probability of absorption for neutrino fields goes to zero as $\omega$ goes to zero. The absorption probability for a mode with frequency slightly smaller than the superradiant limit $\omega \sim (1-\epsilon^2)\omega_{\rm{sl}}$ is approximately $(1/16)$. (for a nearly extremal black hole $M^2 \sim a^2$, $r_+ \sim M$)  The possibility of absorption of the modes $\omega < m\Omega$  confirms  the generic violation of CCC by neutrino fields.  The absorption of these low energy modes is not allowed, and the stability of the event horizon is maintained in the case of bosonic fields.

At this moment, we are unable to estimate the magnitude of gravitational self-force effects for massless fields. It is a safe assumption that self-force effects will be comparable to the terms giving rise to naked singularities due to challenging fields in the frequency range $\omega_{\rm sl} < \omega < \omega_{1}$; and there is no doubt that it will be negligible as $\omega$ is lowered to zero. However the probability of absorption also decreases quadratically to zero for incoming waves in these modes. From that point of view the region of frequencies slightly smaller than $\omega_{\rm sl}$ seems to be optimal for overspinning to occur; both to overcome back-reaction effects and maintain a reasonable probability of absorption. The optimal range could be defined with better precision with a proper estimate of self-force effects.

Previously, the transmission and reflection coefficients for massless fields were interpreted as absorption probabilities for single particles~\cite{richartz,matsas,richartz2}. In \cite{superrad} we argued that the concepts of reflection/transmission coefficients
do not represent probabililities for single particles to
be reflected/absorbed; since the relevant equations allow particle creation. We also showed that the phenomenon of spontaneous emission, which we call the Zel’dovich-Unruh effect, acts as a cosmic censor to completely dominate the effect of a single particle to overspin a black hole. Hod also argued that the absorption probability derived from wave equations was wrong, and substituted it with the conditional probability $p(0|1)$ of one fermion being incident upon a nearly extremal  black hole, with no fermions getting reflected back to infinity~\cite{hod}. The expression for $p(0|1)$ was derived by Bekenstein and Meisels by thermodynamical arguments in the case of a rotating black hole immersed in a radiation bath~\cite{beken}. The definition of $p(0|1)$ is clear; it neither represents the probability that a single particle is absorbed, nor the probability that an incoming wave is absorbed. Hod evaluates  $p(0|1)$ in the limit that the black hole temperature tends to zero, and finds that it vanishes in that limit. A nearly extremal black hole rapidly pushes itself away from extremality by spontaneous emission particles in the modes $w< m \Omega$. Therefore many fermions  are expected to be emitted from the black hole and observed as reflections at infinity; ie. the probability of observing no reflecting  fermions is expected to vanish. If $p(0|1)$  were not zero, it would be possible to overspin a black hole by sending in a single fermion. This would contradict the fact  that spontaneous emission dominates the effect of a single particle by many orders of magnitude.

It is well known that quantum expressions for single or few particles do not apply to fields. Fields represent a large number of particles; however that number is indefinite since particle creation/destruction is allowed. The behaviour of the particles is guided
by the field in a probabilistic way so that the expectation values of physical quantities of interest can be calculated. Considering the scattering of waves, the meanings of the transmission and reflection coefficients are clear; they represent ratios of energies coming back from the black hole and going into it. This can be verified by writing down the
integrals for energy fluxes at infinity and the horizon.
This formalism is sometimes called first quantization. In the case of massless fields the absorption probability of a wave  can be calculated by using the transmission and reflection coefficients of Teukolsky equations, as done by Page.
Therefore, to avoid any confusion we would like to note that Hod's derivation $p(0|1)=0$ only states that the probability that one fermion is sent in to the black hole and no fermion is reflected back vanishes for nearly extremal black holes. It should not be interpreted as the absorption probability of a fermionic field  in the modes $\omega< m \Omega$ is zero.   

The fact that  spontaneous emission acts as a cosmic censor to prevent overspinning of black holes by single particles naturally arises the question how effective is this phenomenon against challenging fields?  Recently, we have also considered the effect of  black hole evaporation  as a cosmic censor and showed that its effect is rather weak against challenging test fields. It cannot prevent the overspinning of a black hole of  mass $M \gtrsim 10^{18}$ g \cite{duztas2}. 

To summarise, neutrino fields with frequencies $\omega< m \Omega$ can be absorbed and lead to overspinning of black holes. Back-reaction effects become negligible as $\omega$ is lowered. Spontaneous emission dominates the effect of tunnelling of a  single (or few ) neutrinos in the dangerous modes, and prevent overspinning; but its effect is negligible against challenging fields.

\section{The stability paradox}
The fact that superradiance does not occur for neutrino fields is also manifest in the connection relation (\ref{conneut}). Note that the right hand side is positive definite, so the left hand side should also be positive. The amplitude of  the outgoing wave ($Z_{\rm{out}}$)  is never larger than that of the incoming wave. For that reason, regardless of the frequency, each neutrino wave incident on the black hole with amplitude   $Y_{\rm{in}}$ is partially absorbed by the black hole, and partially scattered back to infinity as an outgoing wave; i.e. there is a net absorption of each mode. 

The connection relation (\ref{conneut}) can be re-written in the form of a probability relation by defining $\tilde{Z}_{\rm{out}}=\sqrt{2}Z_{\rm{out}}$.
\begin{equation}
\frac{\vert Y_{\rm{hole}}\vert^2}{\vert Y_{\rm{in}}\vert^2}+\frac{\vert \tilde{Z}_{\rm{out}} \vert^2}{\vert Y_{\rm{in}}\vert^2}=1 \label{conneutprob}
\end{equation}
From (\ref{conneutprob}) it follows that; in Kerr space-time the probability that a purely ingoing neutrino wave at past null infinity is absorbed by the black hole is $\vert Y_{\rm{hole}}\vert^2/\vert Y_{\rm{in}}\vert^2$, and the probability that it is scattered back to reach future null infinity is given by $\tilde{Z}_{\rm{out}} \vert^2/\vert Y_{\rm{in}}\vert^2$, where the amplitudes are functions of $(M,a,l,m,\omega)$. 

At this stage we can state the stability paradox for neutrino perturbations of Kerr black holes. From the connection relation for neutrino fields we can both  infer a rigorous proof of mode stability and the possibility of absorption of low energy modes $\omega < m\Omega$, which can render the event horizon unstable. This leads to a paradox, because mode stability of black holes under neutrino perturbations imply the possibility of the formation of a naked singularity, which would be the worst case scenario in the absence of mode stability.

\section{Conclusions}
In this work, we derived the connection relation for asymptotic solutions of massless Dirac fields on Kerr background. We showed that a rigorous proof of mode stability of neutrino perturbations can be derived from this connection relation unlike the case of bosonic fields for which Whiting's general proof of stability is required. The connection relation also implies that all modes including $\omega < m\Omega$ can be absorbed by the black hole, i.e. there is no superradiance. The absorption of these low energy modes can lead to overspinning of the black hole into a naked singularity~\cite{duztas,superrad,toth,natario}. We argued that radiative and gravitational self force effect cannot compensate for the overspinning of black holes by neutrino fields as it does for test bodies~\cite{antij} and it can --in principle-- do for bosonic fields. We also argued that, though the allowed range of frequencies for overspinning to occur is $(0,\omega_1)$; it is a safe assumption that back-reaction effects will compensate for the terms giving rise to naked singularity as $\omega$ approaches $\omega_1$, and  the absorption probability decreases quadratically to zero as $\omega$ is lowered to zero~\cite{page}. The frequencies slightly smaller than $\omega_{\rm{sl}}$ seem to be optimal for overspinning to occur. With a proper estimate of back-reaction effects the interval $(0,\omega_1)$ would be narrowed to some extent, but would not vanish. For that reason, the overspinning of black holes by neutrino fields is more generic and sound compared to the cases of test bodies and bosonic fields. 

We stated that, the fact that both the mode stability of the black hole under neutrino perturbations, and the instability of the event horizon (therefore the instability of the black hole) in the interaction of the black hole with neutrino fields can be derived from the same connection relation, leads to a paradox.

We should note that the argument in this work is restricted to mode stability. The connection relation (\ref{conneut}) cannot provide a proof of linear stability of Kerr black holes under neutrino perturbations; i.e. it is still possible for a general linear perturbation with a finite initial energy to grow in time without bound.  We also note that the conclusion that the mode stability implies event horizon instability does not apply to bosonic fields, and definitely not to test bodies. For bosonic fields, neither the connection relations provide a proof of stability, nor can the overspinning of the black hole be considered generic.

\section*{Acknowledgments}

The author thanks an anonymous referee for valuable suggestions and comments.  This work is partially supported by Bo\u{g}azi\c{c}i University Research Fund, by grant number 7981. 

\end{document}